\documentstyle[epsfig]{mn}

\def\beq{\begin{eqnarray}}
\def\eeq{\end{eqnarray}}
\def\delv{\Delta v}
\def\kms{\, \rm{km}\,  \rm{s}^{-1}}
\def\cm2{{\, \rm{cm}}^{-2}}

\title{What Damped Ly-alpha Systems Tell Us About the 
Radial Distribution of Cold Gas at High Redshift}

\author[Ariyeh H. Maller, Jason X. Prochaska, Rachel S. Somerville
and Joel R. Primack]{Ariyeh H. Maller$^{1,2}$, Jason X. Prochaska$^3$,
Rachel S. Somerville$^{2,4}$ and Joel R. Primack$^{1}$\\
$^1$Physics Department, University of California, Santa Cruz, 
CA 95064\\
$^2$Racah Institute for Physics, The Hebrew University, 
Jerusalem, ISRAEL, 91904\\
$^3$Observatories of the Carnegie Institution of Washington, 
Pasadena CA 91101\\
$^4$Institute of Astronomy, Cambridge University, Cambridge, 
CB3 0HA, UK\\}

\begin{document}
  
\maketitle

\begin{abstract}
We investigate the properties of damped Lyman-$\alpha$ systems (DLAS)
in semi-analytic models, focusing on whether the models can reproduce
the kinematic properties of low-ionization metal lines
described by \nocite{pw:97,pw:98}{Prochaska} \& {Wolfe} (1997b, 1998). We explore
a variety of approaches for modelling the radial distribution of the
cold neutral gas associated with the galaxies in our models, and find
that our results are very sensitive to this ingredient. If we use an
approach based on \nocite{fe:80}{Fall} \& {Efstathiou} (1980), in which the
sizes of the discs are determined by conservation of angular momentum,
we find that the majority of the DLAS correspond to a single galactic
disc. These models generically fail to reproduce the observed
distribution of velocity widths. In alternative 
models in which the gas discs are considerably 
more extended, a
significant fraction of DLAS arise from lines of sight intersecting
multiple gas discs in a common halo. These models produce kinematics
that fit the observational data, and also seem to agree well with 
the results of recent hydrodynamical simulations. Thus we conclude 
that Cold Dark Matter based models of galaxy formation can be 
reconciled with the kinematic data, but only at the expense of the 
standard assumption that DLAS are produced by rotationally supported 
gas discs whose sizes are determined by conservation of angular 
momentum. We suggest that the distribution of cold gas at high 
redshift may be dominated by another process, such as tidal 
streaming due to mergers.
\end{abstract}


\section{Introduction}

This paper is the first in a series of papers that examines the
properties of Damped Lyman-$\alpha$ Systems (DLAS) in the context of
Cold Dark Matter (CDM) based Semi-Analytic Models
(SAMs). Traditionally, DLAS are believed to be the progenitors of
present day spiral galaxies
\nocite{wolfe:95}({Wolfe} 1995) and thus any model of galaxy formation must also
account for their properties. The current wealth of observational data
on DLAS includes their number density, column density distribution,
metallicities, and kinematic properties --- see
\nocite{lwt:95,stor:96,sw:00,pett:94,lu:96,pett:97,
pw:97,pw:98,pw:99,pw:00,wp:00}{Lanzetta}, {Wolfe} \& {Turnshek} (1995); {Storrie-Lombardi}, {Irwin}, \&  {MCMahon} (1996); {Storrie-Lombardi} \& {Wolfe} (2000); {Pettini} {et~al.} (1994); {Lu} {et~al.} (1996); {Pettini} {et~al.} (1997); {Prochaska} \& {Wolfe} (1997b, 1998, 1999, 2000); {Wolfe} \& {Prochaska} (2000). 
These data
potentially provide important constraints on cosmology and theories of
galaxy formation.
Here we especially focus on the new kinematic data.

Previously, the number density of DLAS
has been used to provide constraints on cosmological models
\nocite{mm:94,kc:94,mb:94,klyp:95}({Mo} \& {Miralda-Escude} 1994; {Kauffmann} \& {Charlot} 1994; {Ma} \& {Bertschinger} 1994; {Klypin} {et~al.} 1995). These studies 
assumed a simple correspondance between collapsed dark matter halos
and cold gas to obtain upper limits on the amount of cold gas that
could be present. Gas cooling, star formation, supernovae feedback,
and ionization were neglected. A different approach was used by
\nocite{lwt:95,wlfc:95,pf:95,pfh:99}{Lanzetta} {et~al.} (1995); {Wolfe} {et~al.} (1995); {Pei} \& {Fall} (1995); {Pei}, {Fall} \& {Hauser} (1999), in which the observed 
metallicities and {\emph{observed}} number 
densities of the DLAS were used to model global star formation and
chemical enrichment in a self-consistent way. The latter approach was
set in a classical ``closed-box'' style framework rather than a
cosmological context.

Clearly, in order to model DLAS realistically one needs to include the
astrophysical processes of gas dynamics and cooling, star formation,
and chemical enrichment within a cosmological framework. However, this
is a challenge with our current theoretical and numerical
capability. Cosmological $N$-body simulations with hydrodynamics are
hampered by the usual limitations of volume and resolution. This is
apparent in, for example, the recent work by \nocite{gard:99}{Gardner} {et~al.} (1999), in
which it was found that 
even rather high-resolution hydrodynamical 
simulations could not account for most
of the observed DLAS. \nocite{gard:99}{Gardner} {et~al.} (1999) concluded that the majority 
of the damped Ly-$\alpha$ absorption
must arise from structures below the resolution of
their simulations. In addition, it is well known that such
simulations fail to reproduce the sizes and angular momenta of present
day observed spiral galaxies \nocite{stei:99}({Steinmetz} 1999). One might therefore 
be suspicious of the
accuracy of their representation of the spatial distribution of the
cold gas that gives rise to DLAS at high redshift. Because
observational samples of DLAS are cross-section weighted, these
properties are likely to introduce crucial selection
effects. Semi-analytic approaches can deal with nearly arbitrary
resolution and volumes, but are limited in the sophistication and
accuracy of their physical ``recipes''. In particular, most previous
SAMs have focussed on the bulk properties of galaxies, and have not
attempted to model the spatial location of galaxies relative to one
another or the spatial distribution of gas and stars within galaxies.

The only previous attempt to model the properties of DLAS in a CDM 
framework is the work of \nocite{kauf:96}{Kauffmann} (1996) (hereafter K96).  In K96 
the radial distribution of cold gas in galactic discs was modelled by
assuming that the initial angular momentum of the gas matched that of
the halo, and that angular momentum was conserved during the
collapse. Star formation was then modelled using the empirical law of
\nocite{kenn:89,kenn:98}{Kennicutt} (1989, 1998), in which the star formation is a function
of the surface density of the gas, and cuts off below a critical
threshold density. K96 then showed that the number density, column
density distribution, and metallicities of observed DLAS could be
reasonably well reproduced within the Standard Cold Dark Matter (SCDM)
cosmology, and predicted the distribution of circular velocities of
discs that would give rise to DLAS. Assuming that each observed DLAS
corresponds to a single galactic disc, this can then be compared with
the observed distribution of velocity widths derived from the
kinematics of unsaturated, low-ionization metal lines
\nocite{pw:97,pw:98}({Prochaska} \& {Wolfe} 1997b, 1998).

\nocite{pw:97}{Prochaska} \& {Wolfe} 
found the velocity distribution predicted by K96 
to be strongly inconsistent with their data. Furthermore \nocite{jp:98}{Jedamzik} \& {Prochaska} (1998)
showed that the thick rotating disc model favored by \nocite{pw:97}{Prochaska} \& {Wolfe} (1997b)
could only be reconciled with a finely tuned CDM
model. 
But CDM actually predicts that halos will have much substructure, and
\nocite{hsr:98}{Haehnelt}, {Steinmetz} \&  {Rauch} (1998) found that large $\delv$ velocity profiles consistent 
with those observed by \nocite{pw:97}{Prochaska} \& {Wolfe} are produced in their 
very high-resolution hydrodynamical simulations. These
profiles arose not from the rotation of a single disc, but from lines
of sight intersecting multiple proto-galactic
``clumps''. Subsequently,
\nocite{mm:99}{McDonald} \& {Miralda-Escud\'{e}} (1999) also showed with a simple analytical model that 
DLAS produced by intersection with a few gas clouds could create
kinematics consistent with the observations in a CDM universe.

These results were encouraging but remain somewhat inconclusive. The
hydro simulations do not allow the construction of a statistical,
cross-section selected sample of DLAS, so it is difficult to assess
how typical are the systems that they identified. In addition, these
simulations were restricted to a single cosmology (SCDM), and did not
include star formation or supernovae feedback.
The generic difficulty of hydro simulations in producing reasonable
discs at low redshift has already been noted. Therefore a further
investigation using detailed semi-analytic models is worthwhile.

In the standard CDM picture of galaxy formation
\nocite{wr:78,bfpr:84}(based on {White} \& {Rees} 1978; {Blumenthal} {et~al.} 1984) gas is heated to the virial 
temperature when a halo forms and then cools and falls into the centre
of the halo where it subsequently forms stars.  In SAMs, which include
the hierarchical formation of structure, this process happens numerous
times as halos continually merge and form larger structures.  This
naturally results in halos that may contain many gaseous discs, each
one associated with a sub-halo that prior to merging had been an
independent halo. In this paper, we explore the possibility that such
a scenario can account for the observed kinematics of the DLAS in a
manner analogous to the proto-galactic clumps of
\nocite{hsr:98}{Haehnelt} {et~al.} (1998) and the gas clouds of \nocite{mm:99}{McDonald} \& {Miralda-Escud\'{e}} (1999). 
Here, however, the number densities, gas contents, and metallicities
of these proto-galaxies are determined by the full machinery of the
SAMs, which have been tuned to produce good agreement with the optical
properties of galaxies at low and high redshift. We introduce new
ingredients to describe the kinematics of satellite galaxies within
dark matter halos, and the spatial distribution of cold gas in discs.
We also include a model that is not based on the machinery of the SAMs
to demonstrate that our general conclusions are not overly dependent 
on the specifics of how these processes are handled in the SAMs.

We start with a review of the the observational properties of DLAS
(section \ref{observations}). Next, section \ref{models} gives a brief
description of the ingredients of the SAMs, and describes how we
simulate the observational selection process for DLAS and produce
simulated velocity profiles.  We demonstrate in section
\ref{bad_models} that gaseous discs with
sizes determined by conservation of angular momentum fail to match the
kinematic data, and then in section
\ref{bigdiscs} show that acceptable solutions can be found if the
gaseous discs have a large radial extent.  Section \ref{depend}
examines the sensitivity of our results to a number of model
parameters.  In section \ref{compare} we discuss the properties of the
gas discs in our model and compare them to HI observations of local
spirals and with the results of hydro simulations. 
Lastly we close with some discussion and conclusions.

\section{Observational Properties of DLAS} \label{observations}
DLAS are defined as those absorption systems that have a column
density of neutral hydrogen in excess of $2 \times 10^{20}$ atoms per
square centimeter \nocite{wolfe:86}({Wolfe} {et~al.} 1986). \nocite{pw:96,pw:97j}{Prochaska} \& {Wolfe} (1996, 1997a) found that 
the velocity profiles of low ionization state metal lines 
(Si$^+$, Fe$^+$, Cr$^+$, etc.) trace each other well and therefore 
presumably the kinematics of the cold gas. 

They therefore undertook
to obtain a large sample of the kinematic properties of DLAS as
measured by the associated metal lines and compared them to the 
predictions from a number of models.  
All of the observations were obtained with HIRES \nocite{vogt:92}({Vogt} 1992) on 
the 10m Keck~I telescope. None of the DLAS were chosen with 
{\it a priori} kinematic
information and the metal line profiles were selected 
according to strict criteria including that they not be saturated, 
therefore it is believed that the sample is kinematically unbiased.
We have taken care that our model profiles
match the resolution and signal-to-noise of the observations and 
that they conform to the same profile selection criteria.

\nocite{pw:97}{Prochaska} \& {Wolfe} (1997b,  hereafter PW97) developed four 
statistics to characterize the velocity profiles of the gas, which 
we also use to compare our models to the 
data set of 36 velocity profiles in \nocite{pw:98}{Prochaska} \& {Wolfe} (1998) and \nocite{wp:00}{Wolfe} \& {Prochaska} (2000).
The four statistics as defined in PW97 
are:
\begin{itemize}
\item $\delv$, the velocity interval statistic,
defined as the width containing $90\%$ of the optical depth.
\item $f_{mm}$, the mean-median statistic,  
defined as the distance between the mean and 
the median of the optical depth profile, divided by $\delv /2$.
\item  $f_{edg}$, the edge-leading statistic, 
defined as the distance between the highest peak and the mean, 
divided by $\delv /2$.
\item $f_{2pk}$, the two-peak statistic, 
defined as the distance of the second peak to
the mean. Positive if on the same side of the mean as 
the first peak and negative otherwise.
\end{itemize}
The other observational data that our modeling of DLAS must conform 
to are the differential density distribution $f(N)$ (the number of
absorbers per unit column density per unit absorption distance)
and the distribution of metal abundances. 
The most recent determination of $f(N)$ comes from \nocite{sw:00}{Storrie-Lombardi} \& {Wolfe} (2000). 
The metal abundances in damped systems at high redshift ($z>2$)
have most recently been compiled by \nocite{pett:97}{Pettini} {et~al.} (1997) and 
\nocite{pw:99,pw:00}{Prochaska} \& {Wolfe} (1999, 2000).

\section{Models} \label{models}
\subsection{Semi-Analytic Models} \label{sams}
We use the semi-analytic models developed by the Santa Cruz group
\nocite{some:97,sp:99,spf:00}({Somerville} 1997; {Somerville} \& {Primack} 1999; {Somerville}, {Primack} \&  {Faber} 2000), which are based on the 
general approach pioneered
by \nocite{wf:91}{White} \& {Frenk} (1991), \nocite{kwg:93}{Kauffmann}, {White} \&  {Guiderdoni} (1993) and \nocite{cole:94}{Cole} {et~al.} (1994). Our analysis is
based on the fiducial $\Lambda$CDM model presented in
\nocite{spf:00}{Somerville} {et~al.} (2000,  hereafter SPF), which was shown there
to produce good agreement with many properties of the observed
population of Lyman-break galaxies at redshift $\sim2.5-4$, and the
global evolution with redshift of the star formation density,
metallicity, and cold gas density of the Universe. Below we describe
the aspects of the SAMs most relevant to modeling the DLAS, and refer
the reader to SPF and \nocite{sp:99}{Somerville} \& {Primack} (1999,  hereafter SP),
for further details.

\subsubsection{halos and sub-halos}
The number density of virialized dark matter halos as a function of
mass and redshift is given by an improved Press-Schechter model
\nocite{st:99}({Sheth} \& {Tormen} 1999). The merging history of each dark matter halo at a 
desired output redshift is then determined according to the
prescription of \nocite{sk:99}{Somerville} \& {Kolatt} (1999). As in SP, we assume that halos with
velocity dispersions less than $\sim 30 \kms$ are photoionized and
that the gas within them cannot cool or form stars. This sets the
effective mass resolution of our merger trees. When halos merge, the
central galaxy in the largest progenitor halo becomes the new
central galaxy and all
other halos become ``sub-halos''. These sub-halos are placed at a
distance $f_{mrg} r_{\rm vir}$ from the centre of the new halo, where
$r_{\rm vir}$ is the virial radius of the new halo. We will take
$f_{mrg}$ to be 0.5 as in SP, but will examine the importance of this
parameter in section \ref{depend}. 

After each merger event, the satellite galaxies fall towards the
centre of the halo due to dynamical friction. We calculate the radial
position of each satellite within the halo using the differential
formula
\beq
r_{fric}{{dr_{fric}}\over{dt}}
=-0.42 \epsilon^{0.78}{{Gm_{sat}}\over{V_c}}
\ln{(1+{{m_h}\over{m_{sat}}})}.
\eeq
Here $m_h$ and $m_{sat}$ are the masses of the halo and satellite 
respectively, and $\epsilon$ is a ``circularity'' parameter which 
describes the orbit of the satellite and is drawn from a flat 
distribution between 0.02 and 1 as suggested by N-body simulations
\nocite{nfw:95}({Navarro}, {Frenk} \& {White} 1995).  The halos are assumed to have a singular isothermal
density profile and to be tidally truncated where the density of the
sub-halo is equal to that of its host at its current radius. When a
sub-halo reaches the centre of the host, it is destroyed and the
galaxy contained within it is merged with the central galaxy.

\subsubsection{gas and stars} \label{gas}
In our models, gas can occupy one of two phases, cold or hot. Halos
contain hot gas, which is assumed to be shock-heated to the virial
temperature of the halo and distributed like the dark matter in a
singular isothermal sphere (SIS). After a cooling time $t = t_{\rm
cool}$ has elapsed, gas at a sufficiently high density (corresponding
to the gas within the ``cooling radius'' $r_{cool}$) is assumed to 
cool and condense into a disc. This cold gas then becomes available 
for star formation.

Star formation takes place in both a quiescent and bursting
mode. Quiescent star formation proceeds in all discs whenever gas is
present, according to the expression
\beq
\dot{m}_*={m_{cold}\over{\tau_*}} ,
\eeq
where $m_{cold}$ is the mass in cold gas and $\tau_*$ is an efficiency
factor that is fixed using nearby galaxy properties (see below).  
In the bursting mode,
which takes place following galaxy-galaxy mergers, the efficiency of
star formation is sharply increased for a short amount of time ($\sim$
50--100 Myr). The efficiency and timescale of the starbursts has been
calibrated using the results of hydrodynamical simulations as
described in SPF. The merger rate is determined by the infall of
satellites onto the central galaxy, as described above, and the
collision of satellites with one another according to a modified
mean-free path model (see SP and SPF).

In association with star formation, supernovae may reheat and expell
the cold gas from the disc and/or the halo. We model this using the
disc-halo model of SP, in which the efficiency of the feedback is
larger for galaxies residing in smaller potential wells. These stars
also produce metals, which are mixed with the cold inter-stellar gas,
and may be subsequently ejected and mixed with the hot halo gas, or
ejected into the diffuse extra-halo IGM. Our simple constant-yield,
instantaneous recycling model for chemical enrichment produces
reasonable agreement with observations of metallicities of nearby
galaxies (SP), the redshift evolution of the metallicity of cold gas
implied by observations of DLAS, and the metallicity of the
Lyman-$\alpha$ forest (SPF).

The main free parameters of the model are the star formation
efficiency, $\tau_*$, the supernovae feedback efficiency
$\epsilon_{SN}^0$ and the mass of metals produced per unit mass of
stars, or effective yield, $y$. As described in SP, we set these
parameters so that a ``reference galaxy'' with a rotation velocity of
220 $\kms$ at redshift zero has a luminosity, cold gas mass fraction
and metallicity in agreement with local observations. Good agreement
is then obtained with optical and HI properties of local galaxies
(SP), and optical properties of high redshift galaxies (SPF).

\subsubsection{cosmology}
In this paper our fiducial models are set within a $\Lambda$CDM
cosmology with $\Omega_{\Lambda}=0.7,
\Omega_0=0.3, h=0.7$, corresponding to model $\Lambda$CDM.3 in SP, 
and the fiducial model of SPF. We have presented similar results for a
standard CDM ($\Omega_0=1$) cosmology in
\nocite{mspp:99}{Maller} {et~al.} (1999).  As recent observational results seem to favor a 
cosmological constant \nocite{perl:99}({Perlmutter} {et~al.} 1999) and a flat universe 
\nocite{boom:99}({Melchiorri} {et~al.} 1999)
we feel justified in focusing on only this cosmology. In section 
\ref{depend} we show that our results are not very sensitive to the 
assumed cosmology.

We focus our analysis on halos at an output redshift of $z=3$. We have
also performed an identical analysis on halos at $z=2$ and find no
significant differences, consistent with the kinematic data and column
density distribution $f(N)$, which show little evolution over this
range. We expect to see evolution both in low redshift ($z < 1.5$) 
and very high redshift ($z > 4$) systems, however we will defer 
discussion of this to a future paper.

\subsection{The Spatial Distribution of Cold Gas}
The standard SAMs do not provide us with information on the radial
distribution of gas and stars in the model galaxies. It is reasonable
to assume that the surface density of the cold gas is important in
determining the star formation rate in the gaseous discs, and in this
case the radial distribution of gas should be modelled
self-consistently within the SAMs. This has been done in the models of
K96. However, there are many uncertainties attached to modelling the
structure of the gaseous disc in the initial collapse, and how it may
be modified by mergers, supernovae feedback, and secular
evolution. Therefore here we choose a different approach. The SAMs
described above produce good agreement with the observed $z\sim 3$
luminosity function of Lyman-break galaxies (SPF). The total mass
density of cold gas at this redshift is also in agreement with
estimates derived from observations from DLAS
\nocite{stor:96,sw:00}({Storrie-Lombardi} {et~al.} 1996; {Storrie-Lombardi} \& {Wolfe} 2000). We can therefore ask how this gas must be
distributed relative to these galaxies in order to produce agreement
with an independent set of observations, the kinematic data.

We assume that the vertical profile of the gas is exponential, and
consider two functional forms for the radial profiles of the cold gas:
exponential and $1/R$ (Mestel). The exponential radial profile is
motivated by observations of local spiral galaxies, which indicate
that the light distribution of the disc is well fit by an exponential
\nocite{free:70}({Freeman} 1970). If one assumes that as cold gas is converted into
stars its distribution doesn't change 
(which many theories of disc sizes implicitly assume),
then the profile of cold gas at high
redshift should also be exponential. The column density of the gas may
then be parameterized by two quantities, the scale length $R_g$ and
the central column density $N_0 \equiv m_{\rm gas}/(2\pi
\mu m_H R_g^2)$ (where $m_{\rm gas}$ is the total mass of cold gas 
in the disc, $m_H$ is the mass of the hydrogen atom and $\mu$ is the
mean molecular weight of the gas, which we take to be 1.3 
assuming 25\%
of the gas is Helium). The column
density as a function of radius is given by
\beq
N_{exp}(R)=N_0 \exp{{\left[-{{R}\over{R_g}}\right]}}
\eeq

The $1/R$ profile, sometimes refered to as a Mestel distribution
\nocite{mest:63}({Mestel} 1963), is also motivated by observations. Radio 
observations \nocite{bosma:81}({Bosma} 1981) have shown that the surface density of
HI gas is proportional to the projected surface density of the total
mass, which for a perfectly flat rotation curve would imply a $1/R$
distribution. We parameterize the Mestel disc in terms of the
truncation radius $R_t$ and the column density at that radius $N_t
\equiv m_{\rm gas}/(2\pi \mu m_H R_t^2)$:
\beq
N_{mes}(R)=N_t {{R_t} \over{R}}.
\eeq

In the limit of infinitely thin discs, we can calculate the cross
section for these distributions analytically and use them to check our
numeric code. For an exponential disc, the inclination averaged cross
section is
\begin{eqnarray}
\sigma(N'>N)={{\pi R^2_g \gamma_m^2}\over{2}}
(\ln^2{{N_0}\over{N\gamma_m}}+\ln{{N_0}\over{N\gamma_m}}+\frac{1}{2})
\end{eqnarray}
\nocite{bl:96}({Bartelmann} \& {Loeb} 1996). The variable $\gamma_m=\min[{{N_0}\over{N}},1]$ 
is introduced because a column density of $N$ when $N > N_0$ can only
be reached if the disc is inclined such that $\cos{\theta} >
\gamma_m$. For Mestel discs the corresponding expression is
\begin{eqnarray}
\sigma(N'>N)=\pi R^2_t {N_t\over N}(\frac{1}{2}-\ln{N_t \over N})
\end{eqnarray}
for $N > N_t$.

\subsection{Selecting and Modeling DLAS} \label{dlas}

The fiducial SPF ``standard SAMs'' provide us with a list of galaxies 
contained within a halo of a given mass or circular velocity at a given
redshift. For each of these galaxies, we are also provided with the
internal circular velocity, radial distance from the halo centre,
stellar exponential scale length, and the cold gas, stellar, and metal
content of its disc. We distribute the galaxies randomly on circular
orbits (we discuss the importance of this simplification in section
\ref{depend}) and assign them random inclinations.

We create twenty realizations of a grid of halos with circular
velocities between 50 $\kms$ and 500 $\kms$. These correspond to
different Monte Carlo realizations of the halos' merging histories. We
then choose a model for the radial distribution of the gas and
calculate the surface density distribution, constrained by the total
gas mass as determined by the SAMs. We create twenty random
realizations of the satellite orbits and inclinations in each of the
four hundred halos and calculate the column density along each line of
sight. The number of lines of sight passed through each halo is
determined by the cross-section weighted probability of intersecting a
halo of that mass. 
The total number of lines of sight is chosen to produce about ten
thousand DLAS. Each line of sight that passes through a total column
density exceeding $2 \times 10^{20} \cm2$ is then saved (along with 
all the properties of the halo that it is found in) and analyzed using
the methods of PW97.

To create synthesized spectra we must include substructure in the gas
discs, which we do by assuming that the gas is distributed in small
clouds within the disc. The necessary parameters are: $\sigma_{int}$,
the internal velocity dispersion of each cloud; $N_c$, the number of
clouds; and $\sigma_{cc}$, their isotropic random motions. Following
PW97, we take $\sigma_{int}=4.3 \kms$ and $N_c=5$; both values were
derived from Voigt profile fits to the observations with $N_c=5$ being
the minimum acceptable number of individual components.  
Increasing the cloud
number, $N_c$ to as high as 60 does not improve the goodness of fit
(PW97) for a disc model like we are considering here because our model
discs are relatively thin. Also we take $\sigma_{cc}=10 \kms$, 
since we assume that the gas discs are cold. These internal
velocities are in addition to the circular velocity of the disc and
the motions between discs.  For every line of sight the positions of
the clouds are chosen by taking the continuous density distribution to
be a probability distribution; i.e. the likelihood of a cloud being at
a position in space is proportional to the gas density at that
point. Synthetic metal-line profiles are produced taking into 
account the varying metallicity of the gas in the multiple discs 
along the sightline as given by the SAMs.  The spectrum
is smoothed to the resolution of the HIRES spectrograph
\nocite{vogt:92}({Vogt} 1992), noise is added and then the four statistics of PW97
are applied.  Finally, a Kolmagornov--Smirnoff (KS) test is performed
to ascertain the probability that the data of \nocite{pw:98}{Prochaska} \& {Wolfe} (1998)
and \nocite{wp:00}{Wolfe} \& {Prochaska} (2000) could be a random subset of the model.

It should be noted that while we try to include all of the relevant
physics in the modeling, there are a number of simplifications. The
kinematics of sub-halos within the host halos assumes that the
sub-halos are on circular orbits and utilizes an approximate formula
for the effects of dynamical friction. We assume that the gas discs
have a simple radial profile and are axisymmetric.  Also we assume
that the distribution of the gas does not depend on galaxy environment
or Hubble type. We expect that gas discs should be distorted by the
presence of other galaxies in the same halo or by previous merger
events \nocite{mm:99,kola:99}(cf. {McDonald} \& {Miralda-Escud\'{e}} 1999; {Kolatt} {et~al.} 1999) yet we ignore these effects.  
In the spirit of SAMs we hope
that these assumptions will capture the essential properties of the
resulting DLAS to first order, and we investigate the sensitivity of
our results to some of these assumptions. In section~\ref{compare}, we
note the good agreement of some of the features of our model with the
results of recent hydrodynamical simulations, and in the future we 
hope to refine our modelling by further comparisons with simulations.

\section{Results}
\subsection{Unsuccessful Models: Classical Discs} 
\label{bad_models}
In this section we investigate several models based on standard
theories of the formation of galactic discs. These theories are
generically based on the idea of \nocite{mest:63}{Mestel} (1963) that the specific
angular momentum of the material that forms the galactic disc is
conserved as it cools and condenses.  Since this idea was first
applied in the classic work of \nocite{fe:80}{Fall} \& {Efstathiou} (1980), 
many authors have refined this theory by including the effects of the
adiabatic contraction of the dark halo, the presence of a bulge,
more realistic halo profiles, and disc stability criteria
\nocite{bffp:86,kruit:87,fpbf:93,dss:97,mmw:98,bosch:99}({Blumenthal} {et~al.} 1986; {van der Kruit} 1987; {Flores} {et~al.} 1993; {Dalcanton}, {Spergel} \&  {Summers} 1997; {Mo}, {Mao}, \& {White} 1998; {van den Bosch} 1999). 

In the simplest of such models, we assume a singular isothermal
profile for the dark matter halo, neglect the effects of the halo
contraction on the assembly of the disc, and assume that the profile of
the cold gas after collapse has the form of an exponential. The
exponential scale length is then given by the simple expression:
\beq
R_s={1\over{\sqrt{2}}} \lambda_H r_i \label{eqn:rdisc}
\eeq
where $\lambda_H$ is the dimensionless spin parameter of the halo, and
$r_i$ is the initial radius of the gas before collapse.
In $N$-body simulations, the spin parameter
$\lambda_H$ for dark matter halos is found to have a log-normal
distribution with a mean of about 0.05 \nocite{warr:92}({Warren} {et~al.} 1992). 
A generalization of this model, using an NFW profile for the dark 
matter halo and including the effect of halo contraction, 
has recently been presented by \nocite{mmw:98}{Mo} {et~al.} (1998).

In model EXP1, we assume $\lambda_H = 0.05$ for all halos, take
$r_i = \min[r_{cool}, r_{vir}]$, and calculate the scale length for
each disc from eqn~\ref{eqn:rdisc}. Note that when this approach is
used to model \emph{stellar} scale lengths, the values that we obtain
are in good agreement with observations at redshift zero and redshift
$\sim 3$ (SP; SPF). 

However local observations \nocite{br:97}({Broeils} \& {Rhee} 1997) find that in gas rich 
galaxies the HI disc always has a larger extent than the stellar disc.
To explore this scenario we try a model where the exponential scale
length of the gas is given by a multiple of the stellar disc scale 
length. We find multiplying the scale length calculated from 
eqn.~\ref{eqn:rdisc} by a factor of six (model EXP6)
produces the best agreement with the kinematic data but can still 
be rejected at $> 95\%$ confidence level.

\begin{figure} 
\centerline{\epsfig{file=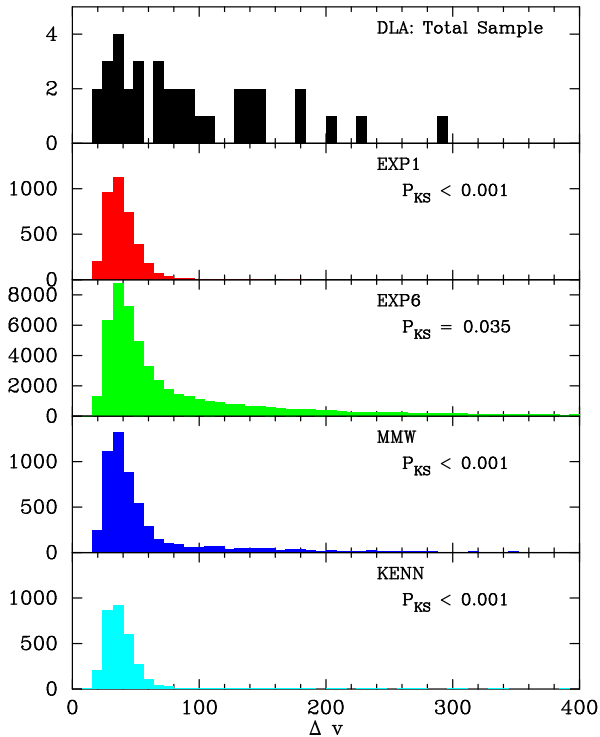, width=\linewidth}}
\caption{The distribution of the $\delv$ statistic from the data
set of \protect\nocite{pw:98}{Prochaska} \& {Wolfe} (1998) and \protect\nocite{wp:00}{Wolfe} \& {Prochaska} (2000) compared to DLAS 
produced in the four models of section~\protect\ref{bad_models}.
}\label{fdvfe}
\end{figure}
In model MMW, we use the fitting formulae of \nocite{mmw:98}{Mo} {et~al.} (1998) to obtain
the scale radius. In this model we do not use the gas content
predicted by the SAMs, but instead, following \nocite{mmw:98}{Mo} {et~al.} (1998) we assume
that the disc mass is a fixed fraction (one tenth) of the total mass
of each halo or sub-halo. 
This procedure produces roughly three times more cold gas 
per halo than the SAMs as there is no star-formation and no hot gas.
Thus this model should be seen
as an upper limit on the amount of cold gas that is available
to form DLAS in the halo mass range we are considering.
The spin parameter $\lambda_H$ is chosen randomly from a log-normal 
distribution,
and the exponential scale length is found from eqn~\ref{eqn:rdisc}.
The main difference between
our MMW model and the actual model of \nocite{mmw:98}{Mo} {et~al.} 
is that we include sub-halos (multiple galaxies in each halo). 
Because \nocite{mmw:98}{Mo} {et~al.} do not simulate the merging history of
their halos, they assume that only one galaxy inhabits each halo
(which would correspond to our central galaxy).

The models of K96 used the assumption that the initial profile of the
cold gas resulted from conservation of angular momentum, and modelled
star formation according to the empirical law proposed by
\nocite{kenn:89}{Kennicutt} (1989). \nocite{kauf:96}{Kauffmann} then found that surface density
of the gas discs tended to remain close to the critical surface 
density, as \nocite{kenn:89}{Kennicutt} in fact observed. 
In our model KENN, based on these observations and
the results of K96, we again take the total mass of cold gas from the
SAMs, and distribute it at the critical density, which for a flat
rotation curve is given by :
\beq \label{eqncr}
N_{cr}=1.5 \times 10^{21}\cm2 \left({{V_c}\over{200\kms}}\right)
\left({{1\, {\rm{kpc}}} \over{R}}\right).
\eeq
Thus for a given $V_c$ this is a Mestel distribution, with $N_t$
determined by the above equation and the total mass of cold gas.

\begin{figure} 
\centerline{\epsfig{file=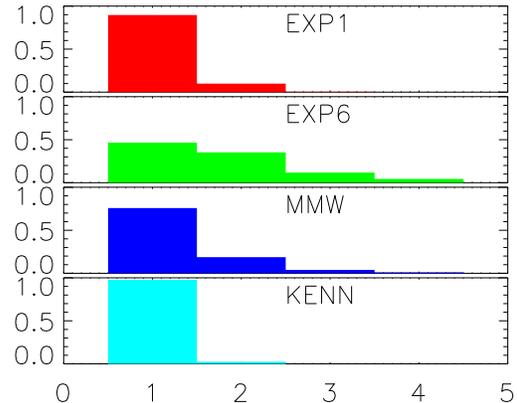,width=\linewidth}}
\caption{Probability distribution of the number of gaseous
discs intersected by a line of sight producing one DLAS. Most of the
DLAS are produced by lines of sight passing through a single disc.
}\label{fnumfe}
\end{figure}
\begin{table}
\begin{tabular}{ccccc}
\hline
Model & $\delv$ & $f_{mm}$ & $f_{2pk}$ & $f_{edg}$ \\
\hline
\hline
EXP1 & $<0.001$ & 0.006 & $<0.001$ & $<0.001$ \\
EXP6 & 0.035 & 0.67 & 0.31 & 0.16 \\
MMW  & $<0.001$ & 0.32 & 0.026 & 0.005 \\
KENN & $<0.001$ & 0.005 & $<.001$ & $.001$ \\
\hline
\end{tabular}
\caption{KS probabilities that the distribution of the four statistics
of PW97 for the observed velocity profiles could have been drawn from
each of the four models of section~\protect\ref{bad_models}. All of
these models can be excluded at high confidence levels.}\label{tbfe}
\end{table}

The KS test results for the four statistics of PW97 for these four
models are shown in Table \ref{tbfe}.  The most important failing of
the models is in the $\delv$ statistic.  Fig.~\ref{fdvfe} shows the
distribution of $\delv$ for the data and models.  The $\delv$ values
produced by these models are peaked around 50 $\kms$ with very few
systems having $\delv > 100 \kms$, in sharp contrast to the data.
This is the same result found in PW97 for a single-disc CDM model
(e.g. the model of K96). It is not surprising, as it turns out that
in these models most DLAS are in fact produced by a single disc, as
shown in Fig. \ref{fnumfe}. Only for the EXP6 model are half of the
DLAS the result of intersections with more than a single gas disc, and
only this model has a $\delv$ distribution that is not rejected at
greater than $99.9\%$ confidence.
\begin{figure*} 
\centerline{\epsfig{file=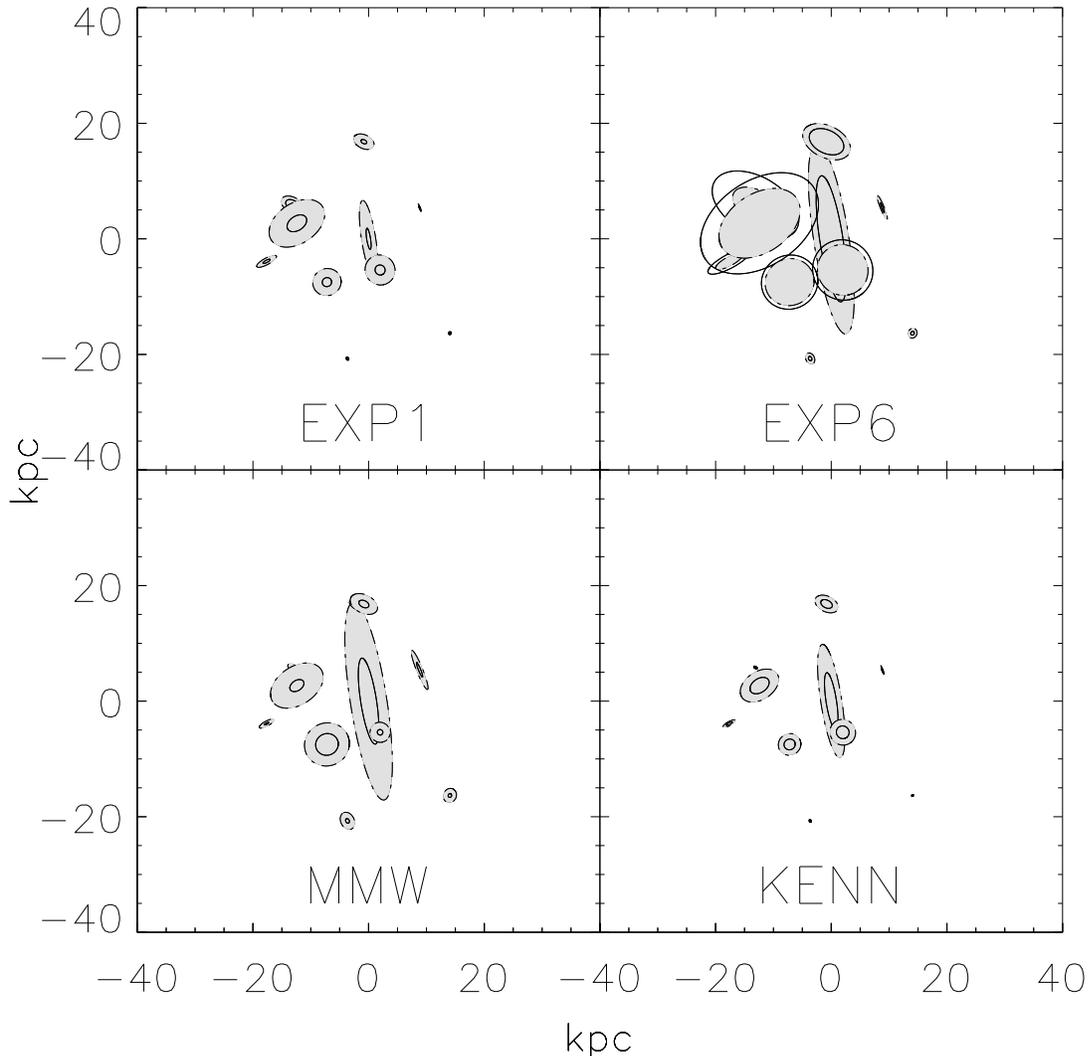, width=\linewidth}}
\vskip 1 cm
\caption{Each panel shows the projection of the gaseous discs in one
example halo of circular velocity 156 $\kms$ for one of the four
models of section~\protect\ref{bad_models}. The shaded region is the
area of the disc with a column density in excess of $2 \times
10^{20}$. The solid line is the half mass radius of the gas disc.  One
can see that in all these models except for EXP6,
a line of sight is unlikely to pass
through more than one gaseous disc. }\label{fdiscs}
\end{figure*}

It is easy to understand why there are so few multiple intersections
in these models by examining Fig. \ref{fdiscs}. This figure shows a
projection of the gas discs residing within a halo of circular
velocity 156 $\kms$.
The sizes of gas discs in these models are much
smaller than the separation between them and thus multiple
intersections are rare.  The sizes of the gas discs in EXP1 and
the KENN model are rather similar.  The gas discs in the MMW model
are generally bigger because there is more cold gas in each disc 
and the log-normally distibuted
$\lambda_H$ varies this compared to the 
EXP1 and KENN models.  In these three models almost all the gas
is above the column density limit to be considered a damped system.
In EXP6 with more extended lower density discs we find some discs 
where a large fraction of their area lies below the damped level.  
More extended exponential discs than those considered here do not 
increase the number of DLAS coming from multiple intersections
because the area dense enough to be above the damped limit rapidly 
shrinks.   

\begin{figure} 
\centerline{\epsfig{file=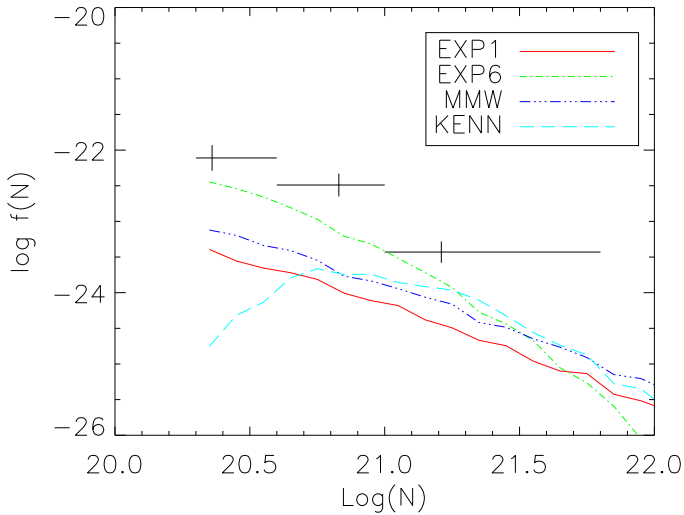,width=\linewidth}}
\vskip .5 cm
\caption{The distribution of column densities,
$\log{f(N)}$ versus $\log(N)$ for the four models of
section~\ref{bad_models} and the data (crosses).  Only the EXP6 model
comes close to fitting the data. The other three models produce too
few DLAS, at least in the $\Lambda$CDM cosmology we are considering.
The KENN model turns over at low column densities because $\log(N_t)$
is often greater than 20.3 (i.e. we ``run out'' of gas before reaching
the cutoff column density defining a DLAS).}\label{ffnfe}
\end{figure}

In Fig.~\ref{ffnfe} we show the column density distribution
$f(N)$ for these models in comparison with the data of
\nocite{sw:00}{Storrie-Lombardi} \& {Wolfe} (2000). Once again, 
of the four models only EXP6 comes close to fitting the data. Thus
although the \emph{total} mass of cold gas is in agreement with that
derived from the observations, the total \emph{cross-section} for
damped absorption is too small if the gas and stars have a similar
radial extent and distribution, as predicted by standard models of
disc formation. We therefore conclude that we may need to consider a
radically different picture of gaseous discs at high redshift.

\subsection{Successful Models: Gas Discs with Large Radial Extent} 
\label{bigdiscs}

In the previous section we found that models in which the sizes of gas
discs at high redshift were calculated from angular momentum
conservation fail to reproduce the kinematics and column density
distribution of observed DLAS. A model based on the observations of
\nocite{kenn:89}{Kennicutt} (1989) for local gas discs and the results of the model of 
K96 also failed. We noted that a common feature of these models is 
that the
majority of DLAS arise from a single galactic disc because of the
small radial extent of these discs compared to their separation. If we
wish to investigate a scenario like the one proposed by
\nocite{hsr:98}{Haehnelt} {et~al.} (1998), in which the kinematics of DLAS arise from lines of 
sight intersecting multiple objects, it is clear that the gaseous
discs must be much larger in radial extent.

\begin{table}
\begin{tabular}{ccccc}
\hline
$\log(N_t)$ & $\delv$ & $f_{mm}$ & $f_{2pk}$ & $f_{edg}$ \\
\hline
\hline
19.3  & 0.04     & 0.01   & 0.14    & 0.27 \\
19.5  & 0.29     & 0.03   & 0.33    & 0.41 \\
19.6  & 0.58     & 0.11   & 0.68    & 0.58 \\
19.7  & 0.18     & 0.29   & 0.53    & 0.73 \\
\hline
\end{tabular}
\caption{KS probabilities for Mestel discs truncated at a 
column density $N_t$. \label{tbbd}}
\end{table}
Unfortunately there does not exist an alternative theoretical
framework for the sizes of gas discs, especially at high redshift, so
in this section we will simply develop a toy model for the
distribution of cold gas.  We hope that the insight gained from such a
toy model will lead to a more physically motivated model in the
future. In our toy models we assume a Mestel profile and assume that
the HI discs are truncated at a fixed column density, perhaps by a
cosmic ionizing background. We investigate a range of values for this
critical column density $N_t$, which is the only additional free
parameter of the model. We take the vertical scale height of the gas
to be half the {\it stellar} disc scale length, as calculated from
eqn.~\ref{eqn:rdisc}. Since the radial extent of the gas is now so
large compared to the stars, this still results in gaseous discs that
are quite ``thin''. We find that our results are only
modestly dependent on the assumed vertical scale height, as we show in
section~\ref{depend} \nocite{mspp:00}(see also {Maller} {et~al.} 2000b). 

\begin{figure} 
\centerline{\epsfig{file=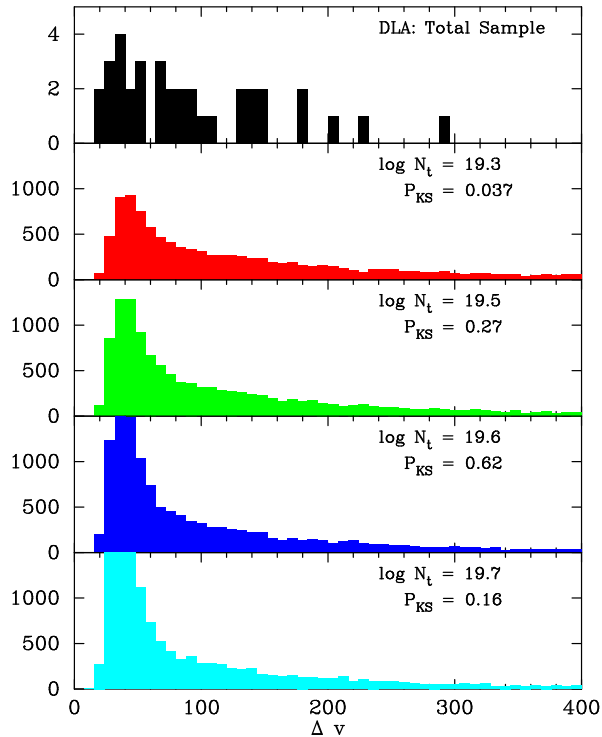,width=\linewidth}}
\caption{The distribution of $\delv$ for the data and four models
where the gas is truncated at a value of
$\log{N_t}=19.3,19.5,19.6,19.7$.  The lower the value of $N_t$, the
larger the radial extent of the gas discs, and therefore the
greater the fraction of DLAS that arise from multiple intersections.
Multiple intersections tend to create DLAS with large $\delv$.
}\label{fdvbd}
\end{figure}

\begin{figure} 
\centerline{\epsfig{file=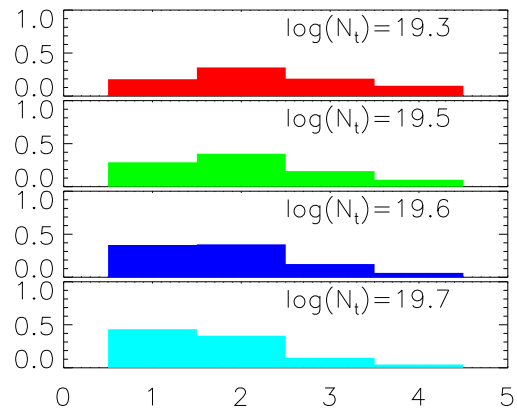,width=\linewidth}}
\caption{The distribution of the number of gas discs that produce
a DLAS for four values of $N_t$. The distribution of $\delv$ is
principally determined by the fraction of DLAS that are produced by
``multiple hits'' (lines of sight that pass through more than one
disc). In this class of models, the best fit to the kinematic data
occurs if $\sim 40$ percent of DLAS come from single discs.
}\label{fnumbd}
\end{figure}

The distribution of $\delv$ for several values of $N_t$ between 2 and
5 $\times 10^{19} \cm2$ are shown in Fig. \ref{fdvbd}. We find that a
value for the truncation column density $N_t$ of $4 \times 10^{19}
\cm2$ (i.e., $\log{N_t}=19.6$)
gives the best fit to the kinematic data.  The distribution now
shows a significant tail to large values of $\delv$, in much better
agreement with the data. For values of $N_t$ less than $4 \times
10^{19} \cm2$ the models produce more large values of $\delv$ than are
seen in the data, while higher values of $N_t$ produce fewer large
values of $\delv$.
Fig. \ref{fnumbd} shows the relationship between the number of gaseous
discs that produce a DLAS and the truncation level $N_t$.  We see that
when $\sim 40$ percent of the DLAS come from a single gas disc we get
the best fit to the kinematic data.  

\begin{figure} 
\centerline{\epsfig{file=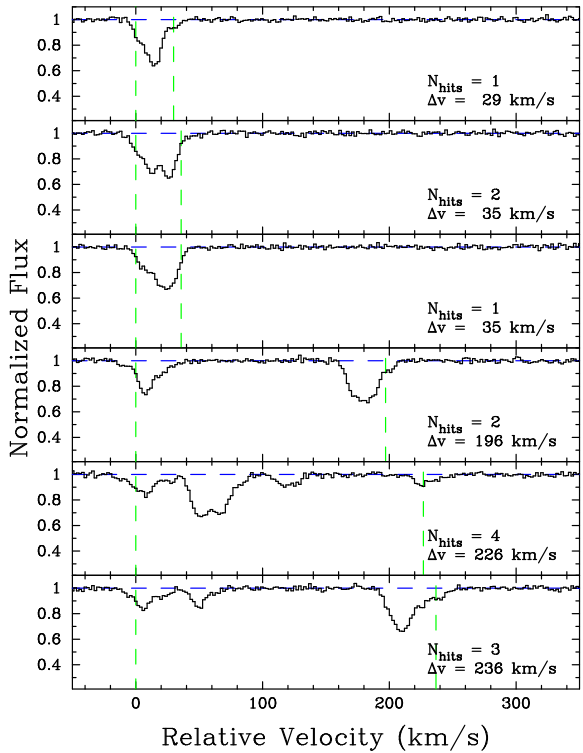,width=\linewidth}}
\caption{Some examples of the velocity profiles of the low 
ions in DLAS produced by our modeling in Section \ref{bigdiscs}.
The dashed line denotes the velocity width $\delv$ of the profile.
The profiles with $\delv >100 \kms$ have large parts of velocity 
space with no absorbtion, something not seen in the data.  
It is possible that this is an artifact of the simplicity of our 
modeling.
}\label{fprofil}
\end{figure}

Table \ref{tbbd} shows the KS
probabilities for the four statistics of PW97. The mean-median
statistic ($f_{mm}$) shows a clear trend with $N_t$, such that the
statistic improves for higher values of $N_t$.  This is because
multiple intersections can produce values of $f_{mm}$ between 0.8 and
1, which are not found in the data. These high values occur in
velocity profiles of two narrow peaks separated by a large ``valley''.
An example of this type of profile is the fourth system in
Fig. ~\ref{fprofil}.  While the statistics of PW97 show agreement 
between this model and the data, the profiles with $\delv > 100$~km/s,
whose kinematics are dominated by the motions of the multiple discs 
relative to one another, show large parts of velocity space with 
no absorbtion (Fig. ~\ref{fprofil}).  
This is something which is not seen in the data.
It is possible that such profiles arise because of the simplicity of 
our modeling and that in a more physical scenario this configuration 
would not occur.

\begin{figure} 
\centerline{\epsfig{file=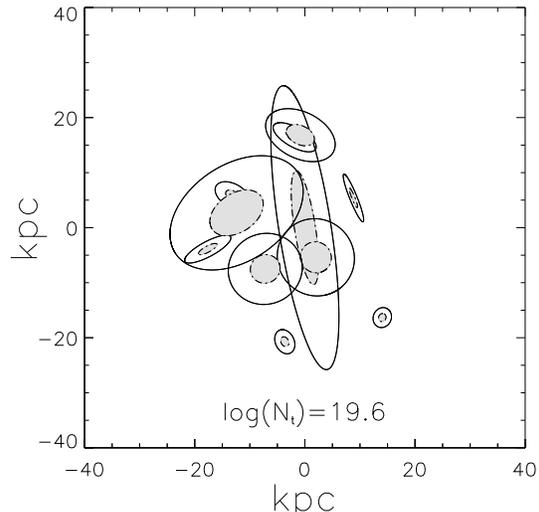,width=\linewidth}}
\vskip .5 cm
\caption{The projection of the gaseous discs in one
example halo of circular velocity 156 $\kms$ for the $N_t=4 \times
10^{19} \cm2$ model. The shaded
region is the area of the disc with a column density in excess of 
$2 \times 10^{20}$ and the solid line is the half mass radius of the 
gaseous disc. One clearly sees that this model will give rise to many
lines of sight with multiple intersections. 
}\label{fbd}
\end{figure}

The gas discs have such a large radial extent in this model (see Fig.
\ref{fbd}) that they will clearly be perturbed by one another and
not retain the simple circular symmetry that we are imposing. Perhaps
a model in which most of the gas is in tidal streams would be more
appropriate. Or perhaps the cold gas is not associated with the
individual galaxies at all, but then we must understand what keeps 
it from being ionized by the extra-galactic UV background. 
Our toy model demonstrates that
the cold gas must somehow be distributed with a very large covering
factor in order to reproduce the observed kinematics of the DLAS;
understanding how it attains this distribution will require furthur 
study and most likely hydro simulations.

\begin{figure} 
\centerline{\epsfig{file=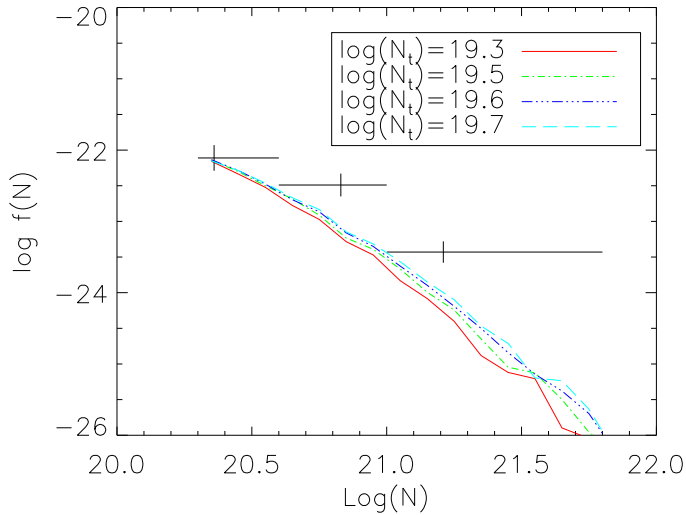,width=\linewidth}}
\vskip .5 cm
\caption{The differential density distribution, $f(N)$ of damped
absorbers from the models (lines) compared to the data of
\protect\nocite{sw:00}{Storrie-Lombardi} \& {Wolfe} (2000) (crosses). These models fit the two lower
column density bins well, but falls short of the highest column
density bin.  }\label{ffnbd}
\end{figure}

We now investigate the column density distribution $f(N)$ and the
metal abundances for these models.  Fig. \ref{ffnbd} shows $f(N)$ for
the models and the data. The column density distribution is not very
sensitive to the truncation density $N_t$, and all the models produce
about the right number of absorbers except in the highest column
density bin. This may be due to our simplistic assumption that all gas
discs are truncated at the same column density.  If a small fraction
of them were much denser it might be possible to have enough high
column density systems without significantly affecting the kinematic
properties of the absorbers. It should also be noted that the data was
tabulated assuming a $q_0=0$ cosmology, which may introduce an
additional discrepancy in comparing to our $q_0=-0.55$ $\Lambda$CDM 
cosmology. We note that the shape of the distribution is
fairly similar to that of the data.

\begin{figure} 
\centerline{\epsfig{file=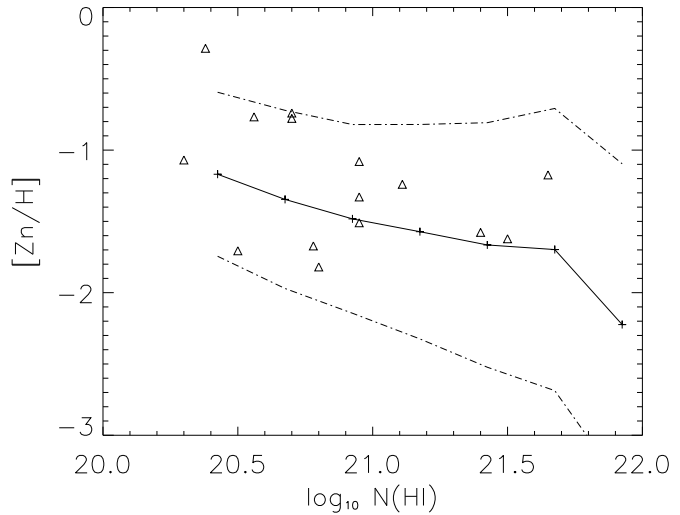,width=\linewidth}}
\vskip .5 cm
\caption{Log(Zn/H) compared to the solar value versus HI column 
density.  The solid line is the average value from our best-fit model
and the dot-dash line shows the one sigma spread.  The data 
(triangles) is from \protect\nocite{pett:97,pw:99}{Pettini} {et~al.} (1997); {Prochaska} \& {Wolfe} (1999).  
The spread and trend
with column density of the data are consistent with that of our model.
}\label{fmetal}
\end{figure}

We also show the average metal abundances of our absorbers versus HI
column density in Fig. \ref{fmetal}. The observational data points 
are [Zn/H] measurements of DLAS with $z>2$ \nocite{pett:97,pw:99}({Pettini} {et~al.} 1997; {Prochaska} \& {Wolfe} 1999).
One can see our model gives
DLAS with metallicities in agreement with the data, and
also reproduces the observed trend with HI column density.
One might expect that systems with higher HI column
densities should have higher metallicities because they are more 
likely to be in more massive halos.  However, because we truncate 
all gas discs at the same value of $N_t$ 
the distribution of column densities 
is the same for all halos masses.  Thus our parameterization 
naturally explains the flat distribution of metal abundances with HI
column densities.

\section{Model Dependencies} \label{depend}
We have presented a model that can produce the observed kinematic
properties of the DLAS as well as the other known properties of these
systems. In this section we examine the sensitivity of our model to
some of the simplifying assumptions we have made.  We examine the
effect of changing the disc thickness, the orbits of the satellites,
the cosmological model, and the assumption of rotationally supported
discs.  We find that none of these have a large effect on the
kinematics of the DLAS in our models.  Table~\ref{tbdep} shows the KS
probabilities when these various assumptions are changed.  The effect
of any of these changes is less than changing the truncation density
from $N_t = 4 \times 10^{19} \cm2$ to $5 \times 10^{19} \cm2$.  We
conclude that our general conclusions are not sensitive to any of
these assumptions.

\subsection{Disc Thickness}
In section \ref{bigdiscs} we assumed the vertical scale length of
the gas to be one half the {\it stellar} disc scale length. Because the gas
is so much more radially extended than the stars, this still resulted
in very thin discs. One might think that as long as $h_z$ is small
compared to the radial size of the discs its exact value would not be
important.  However, as explained in
\nocite{mspp:00}{Maller} {et~al.} (2000b) very thin discs have an increased cross section to
being nearly edge-on, which changes their kinematic properties. Thus
the KS probabilities change non-trivially when we consider thinner
discs with $h_z=0.1R_*$. We favour the model with $h_z=0.5R_*$ because
these large discs are very likely to be warped by interactions, and
\nocite{pw:98}{Prochaska} \& {Wolfe} (1998) have shown that using a larger scale height has an 
effect similar to including warps in the discs. Increasing the disc 
thickness to $h_z=R_*$ also has a non-negligible effect on the 
kinematics because thicker discs create a larger $\delv$ for a 
single disc encounter. Thus there is a trade-off, and we see that we 
can reproduce the kinematics either with thinner discs with larger 
radial extent, or thicker discs with smaller radial extent.

\subsection{Circular Orbits}
We have assumed that all the satellites are on circular orbits within
the halo, which is clearly unrealistic. To test the importance of this
assumption we explore the opposite extreme, which is to assume that
all satellites are on radial orbits.  The potential of a SIS is
$\Phi(r)=V^2_c \ln(r)$ so conservation of energy gives us
\beq
v(r)=\sqrt{2}V_c \sqrt{\ln{r_m/r}}
\eeq
where $r_m$ is the maximum radius the satellite reaches. From this
one can compute that the time it takes for a satellite to travel 
from $r_m$ to $r$ is given by
\beq
t(r)={{r_m}\over{\sqrt{2}V_c}}{\rm{erf}}(\ln{r_m/r}).
\eeq
and that the orbital period is $P=2\sqrt{2\pi}{{r_m}/{V_c}}$.  We find
that the expression $r(t)=r_m(1-.75(4t/P))$ for $0<t<P/4$ is a
reasonable fit to the true function (satellites spend less than 5\% of
their time in the inner fourth of the orbit). We therefore use it to
determine the probability distribution of satellites along their
radial orbits.  From Table~\ref{tbdep} we see that assuming all radial
orbits slightly improves the statistics of our model and thus
considering a true distribution of orbits will probably only increase
the agreement between the data and our model, but not enough to rescue
any of the unsuccessful models.

\subsection{The Initial Infall Radius of Satellites}
As explained in section \ref{sams}, when halos merge the satellite
galaxies are placed at a distance $f_{mrg}$, in units of the virial
radius, from the centre of the new halo.  One might worry that this
parameter, by influencing the position of satellite galaxies in the
halo, may be crucial to our DLAS modeling.  To test this we try a
model with $f_{mrg}$ set to 1.0 (instead of the 0.5 as it has been up
to now). This requires us to change the free parameters of the SAMs to
maintain the normalization of the reference galaxy, as described in
SP. The results of this model are shown in Table
\ref{tbdep}. Doubling this parameter results in only a modest change 
in the kinematic properties of DLAS.
This is because the important factor is the number of galaxies in 
the inner part of the halo (which will give rise to multiple 
intersections); satellites that start their infall from farther out 
still spend a similar amount of time near the central object 
which explains the modest effect on the kinematic properties.
\begin{table}
\begin{tabular}{ccccc}
\hline
Model & $\delv$ & $f_{mm}$ & $f_{2pk}$ & $f_{edg}$ \\
\hline
\hline
Default Model & 0.57 & 0.13 & 0.73 & 0.58 \\
$h_z=0.1R_*$  & 0.22 & 0.12 & 0.57 & 0.34 \\
$h_z=1 R_*$   & 0.28 & 0.10 & 0.30 &  0.46 \\ 
Radial Orbits & 0.60  & 0.29 & 0.75 & 0.66 \\
$f_{mrg}=1$ & 0.36 & 0.14 & 0.61 & 0.63 \\
$\Lambda$CDM.5 & 0.72 & 0.10 & 0.50 & 0.64 \\
OCDM  & 0.28 & 0.12 & 0.66 & 0.68 \\
$z=2$ &	0.45 & 0.17 & 0.61 & 0.67  \\
Nonrotating Discs & 0.35 & 0.05 & 0.42 & 0.34  \\
\hline
\end{tabular}
\caption{KS probabilities for variations of the fiducial model.
}\label{tbdep}
\end{table}
\subsection{Cosmology}

So far we have only considered models set within the currently
favoured $\Lambda$CDM.3 cosmology. However, we would like to know how
sensitive our results are to the assumed cosmology. We consider two
other cosmologies, a flat universe with $\Omega_0=0.5$
($\Lambda$CDM.5) and an open universe with $\Omega_0=0.3$ (OCDM.3; as
in SP). The free parameters must be readjusted for each cosmology as
described in SP. The KS probabilities are listed in Table
\ref{tbdep}.  One sees that the effect of changing the cosmological
model is not a drastic one. We do not show the results here, but we
note that our conclusions concerning the dynamical tests
do not change even in a cosmology with very
low $\Omega_0=0.1$, nor do they change if we assume $\Omega_0=1$.
The total mass of cold gas as a function of redshift is rather 
sensitive to cosmology, however the distribution of $\delv$ is 
almost completely insensitive to cosmology. This is because the 
distribution of $\delv$ from single hits depends only on the 
\emph{shape} of the power spectrum, which is very similar on these 
scales for any CDM model. The contribution from multiple hits is 
determined by the dependence of the merger rate on halo mass, 
which again is a weak function of cosmology. 

\subsection{Non-Rotating Gas}
The final assumption we explore is seemingly a key one: that the cold
gas is rotationally supported in discs. We consider a simple 
alternative model where the cold gas has a bulk velocity with
the same magnitude as the circular velocity of the halo, but in a 
random direction. This might
represent gas that is dominated by streaming motion or infall
rather than rotation. We are still able to reproduce the observed 
kinematics, because they are dominated by the motions of the various 
sub-halos not the motions within the discs.  Thus 
the fundamental assumption that 
cold gas at high redshift is in rotationally supported discs may 
need to be reconsidered.

\section{Properties of Gaseous Discs in our Model} \label{compare}
We have demonstrated that the standard theories of disc formation
cannot reproduce the observed properties of DLAS, and have proposed a
rather unorthodox alternative which succeeds in reproducing these
observations. Here we compare our models with observations of local
discs and with results from recent hydrodynamical simulations to
assess whether the model is reasonable. These results are all for the
fiducial model of section~\ref{bigdiscs}.

Because local gas discs do not show a common surface density profile,
it is common practice to cite their properties out to some surface
density contour, often taken to be 1 $M_{\sun}$pc$^{-2}$ which is 
equal to $1.25 \times 10^{20} \cm2$. We will denote the radius where 
the column density reaches this value as $R_{\rm HI}$.  One observed 
local property of gas discs is that the average surface density
$<\sigma_{\rm HI}>$ out to this level is approximately constant with a
value $3.8 \pm 1.1 M_{\sun}$pc$^{-2}$ \nocite{br:97}({Broeils} \& {Rhee} 1997). Because in our
model all galaxies are normalized by the same value of $N_t$, the
average surface density is identically equal to $2 M_{\sun}$pc$^{-2}$.
The galaxies in our model (at $z \sim 3$) thus have an average surface
density half the local value and share the property that this value is
independent of gas mass.

\begin{figure} 
\centerline{\epsfig{file=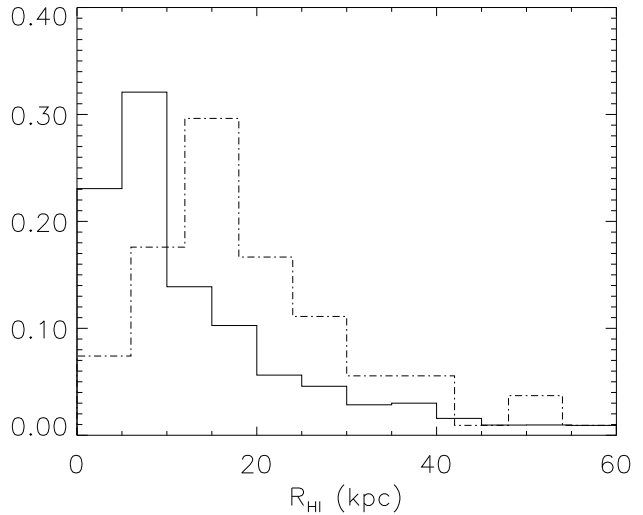,width=\linewidth}}
\vskip .5 cm
\caption{a) The distribution of $R_{HI}$ is plotted for gas discs 
giving rise to DLAS at $z=3$ in our fiducial model (solid) and for the
data of \protect\nocite{br:97}{Broeils} \& {Rhee} (1997)
(dashed). }\label{fsizes}
\end{figure}

We can also compare the sizes of gaseous discs in our models to the
observations of \nocite{br:97}{Broeils} \& {Rhee} (1997).  Fig. \ref{fsizes} shows the
distribution of $R_{\rm HI}$ for the discs that give rise to DLAS in
our model (at $z=3$) and for the local data.  The local discs are
about twice as large as the high-redshift discs producing the DLAS in
our model.  The gas discs of the model however extend another factor
of 3 before they are truncated, something which is not usually seen
locally. Note that as these populations are selected in very different
ways, in addition to being at very different redshifts, it is not
clear that the distributions should agree closely. However, we see
that the radial extent of the gas in our model is not that drastically
different from that in local spiral galaxies.

Local HI surveys find no systems with average surface densities 
less than $5 \times 10^{19} \cm2$ \nocite{zwaan:97}({Zwaan} {et~al.} 1997)
which is attributed to photo-ionization of HI discs below a column 
density of a few $10^{19} \cm2$ by the extra-galactic UV background
\nocite{cs:93,malo:93}({Corbelli} \& {Salpeter} 1993; {Maloney} 1993).
Thus the value of $N_t$ that we attain from considerations of the
DLAS kinematics and number density is surprisingly close 
to local estimates.

If the gas discs in our toy model are truncated because of 
photo-ionization then we would expect the gas near the truncation
edge to have a high ionization fraction.  However this does not 
translate into DLAS with high ionization fractions because this low
column density gas will only be a fraction of the gas that composes 
a DLAS.  Most of the column density of a DLAS will come from gas at
higher densities, as the total needs to be in excess of $2 \times 
10^{20} \cm2$, and thus the average ionization state of the gas 
will be low, in agreement with the observations.  This model would 
predict that the lower column density components of the velocity 
profile would be more likely to have higher ionization states, 
something that can be checked in the existing data.

\begin{figure} 
\centerline{\epsfig{file=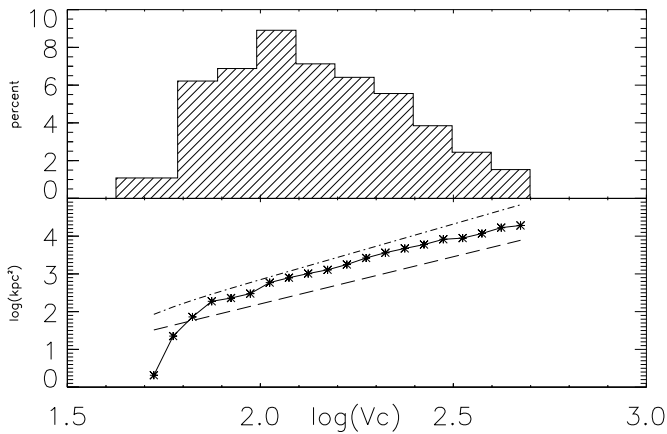,width=\linewidth}}
\caption{The top panel shows the distribution of circular velocities
of the halos that give rise to DLAS in our models.  The bottom panel
shows the average cross section to damped absorption as a function of
circular velocity from our model (solid) and according to 
\protect\nocite{gard:97}{Gardner} {et~al.} (1997) (dot-dash) and  
\protect\nocite{hsr:99}{Haehnelt} {et~al.} (1999) (dashed). However, \protect\nocite{gard:99}{Gardner} {et~al.} (1999) 
finds a much shallower slope of 0.9.
}\label{fvc}
\end{figure}

\begin{figure} 
\centerline{\epsfig{file=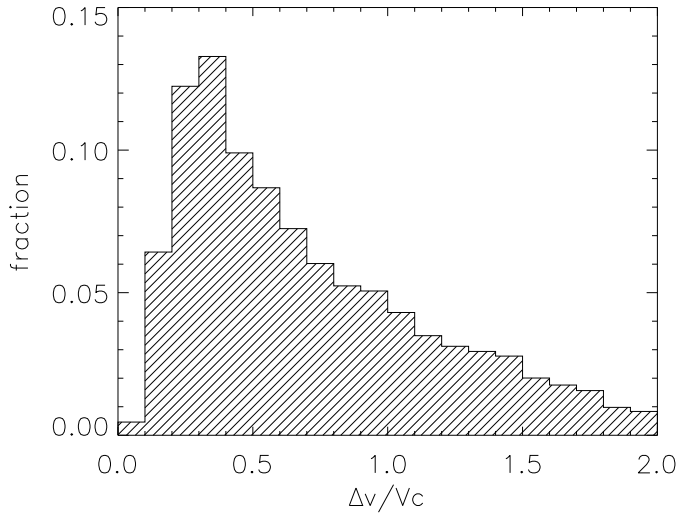,width=\linewidth}}
\vskip .5 cm
\caption{The relationship between the velocity width observed and
the circular velocity of the halo in our toy model.  
}\label{fdvvc}
\end{figure}

It is also interesting to investigate the distribution of halo masses
giving rise to DLAS.  Fig. \ref{fvc} shows the distribution of
circular velocities of the halos containing discs that give rise to
DLAS.  Also shown is the average cross section for DLAS as a
function of circular velocity, which agrees fairly well with the
results of \nocite{gard:97}{Gardner} {et~al.} (1997) (slope = 2.94) and
\nocite{hsr:99}{Haehnelt}, {Steinmetz} \&  {Rauch} (1999) (slope = 2.5), but not those of \nocite{gard:99}{Gardner} {et~al.} (1999)  
who finds a much shallower slope of 0.9. \nocite{hsr:99}{Haehnelt} {et~al.} determine
their average cross section by fitting 
to the observed $\delv$ distribution so we expect that 
the relationship between the circular velocity of the halo 
and the $\delv$ of the DLAS that arise in it must be the same
in our modeling and the simulations of \nocite{hsr:99}{Haehnelt} {et~al.}~. This is
in fact the case as can be seen by comparing from 
Fig. \ref{fdvvc} and Fig. 1 in \nocite{hsr:99}{Haehnelt} {et~al.} (1999).
This seems to suggest that the very different 
approaches of hydro simulations and SAMs are converging on a common
picture for the nature of the DLAS.

\section{Discussion and Conclusions} \label{conclusions}

We have explored the properties of DLAS in semi-analytic models of
galaxy formation. These models produce good agreement with many
optical properties of galaxies at low and high redshift, and the total
mass of cold gas at redshift $\sim 3$ is also in reasonable agreement
with observations. It is therefore interesting to ask whether the
kinematic properties, metallicities, and column densities of DLAS in
these models are in agreement with observations. We investigated the
dependences of these properties on cosmology, the distribution of
satellite orbits, and gaseous disc scale height,
and found that our results were not
sensitive to these assumptions. Our results are \emph{extremely 
sensitive} to our assumptions about the radial distribution of cold
gas within galactic discs. Given that one believes the other
components of our model, one can then perhaps learn about the
distribution of cold neutral gas at high redshift.

Currently popular theories of disc formation posit that the radial
size of a galactic disc is determined by the initial specific angular
momentum of the dark matter halo in which it forms, and that the cold
gas traces the stellar component. Often, the profile of the disc is
assumed to have an exponential form. We investigate several variants
of such models, based on ideas in the literature such as
\nocite{fe:80}{Fall} \& {Efstathiou} (1980), \nocite{mmw:98}{Mo} {et~al.} (1998) and \nocite{kauf:96}{Kauffmann} (1996). We find that the 
kinematics of DLAS arising in such models are in strong conflict with
the observations of \nocite{pw:97,pw:98}{Prochaska} \& {Wolfe} (1997b, 1998). This is consistent with the
previous work of \nocite{pw:97}{Prochaska} \& {Wolfe}, in which it was shown that if
the $\delv$ of each DLAS arises from a single rotating disc, generic
CDM models can be ruled out at high confidence level. Our work has
shown that in theories of disc formation based on angular momentum
conservation, the resulting gaseous discs are so small that most DLAS
are produced by a single disc and thus the models suffer from the
familiar difficulties with the observed kinematics. In addition,
although the total mass of cold gas in the models is in agreement with
the estimate of $\Omega_{\rm gas}$ from \nocite{stor:96}{Storrie-Lombardi} {et~al.} (1996),
when we use realistic cross-section-weighted column-density criteria
to select DLAS in our models, we find that the overall number density
of damped systems is too small. This again seems to indicate that the
covering factor of the gas is too small.

We therefore abandon the standard picture of discs and
investigate a toy model in which the gas is distributed according to a
Mestel distribution with a fixed truncation radius. We adjust the
truncation radius as a free parameter to find the best fit with the
observations, and find that the best-fit value is consistent with the
expected ionization edge due to a cosmic ionizing background. This
results in gas discs which are considerably more radially extended
than the standard models discussed above. For this class of models, we
find good agreement with the four diagnostic statistics of PW97 which
describe the kinematics; in particular, the distribution of velocity
widths $\delv$ in the models now has a tail to large $\delv \sim
200-300 \kms$ as in the observations. In the models, the majority of
these large $\delv$ systems arise from ``multiple hits'', lines of
sight that pass through more than one rotating disc, as in the picture
proposed by \nocite{hsr:98}{Haehnelt} {et~al.} (1998). The column density distribution and
metallicities of the DLAS in the models are also in reasonable
agreement with the observations.

This working model for DLAS has many additional implications that 
may be tested by observations in the near future.  One interesting 
issue is the relationship between DLAS and the Lyman-break galaxies
\nocite{stei:96,lowe:97,stei:98}({Steidel} {et~al.} 1996; {Lowenthal} {et~al.} 1997; {Steidel} {et~al.} 1998), about which little is currently known
observationally \nocite{djor:97,mw:98}({Djorgovski} 1997; {Moller} \& {Warren} 1998).  
Previous theoretical predictions used simplifed relations to 
estimate luminosities \nocite{mmw:98,hsr:99}({Mo} {et~al.} 1998; {Haehnelt} {et~al.} 1999).
Because the SAMs include detailed modelling 
of star formation-related processes as well as full stellar 
population synthesis, we are in a position to make much more detailed
and perhaps more reliable predictions.
Our model suggests that $20\%$ of DLAS contain at least one 
galaxy with an $R$ magnitude brighter then 25.5 in the same dark 
matter halo.  The median projected distance between the DLAS
and the Lyman-break galaxy is about 30 kpc \nocite{mpsp:00}({Maller} {et~al.} 2000a).

Another interesting comparison is with the kinematics of the high
ionization state elements \nocite{wp:00}({Wolfe} \& {Prochaska} 2000).  In the simple picture of the
SAMs, these profiles would naturally be associated with the gas that 
has been shock heated to the virial temperature of the halo.  The hot
gas is distributed spherically in the sub-halo, unlike the cold gas,
which would explain why the velocity profiles of the high ions do
not trace the low ions \nocite{pw:97j}({Prochaska} \& {Wolfe} 1997a).  However, the velocity widths 
of the two profiles, which are dominated by the motions between 
sub-halos within the same larger halo, would be related.

The kinematics and $f(N)$ distribution for absorbers below the damped
limit but with column densities above the value where the disc is
truncated (i.e. Lyman-limit systems) also provide an interesting 
test of our model.  Our modeling suggests that the incidence of 
these absorbers arising from cold discs should increase with the 
same slope as in Fig.~\ref{ffnbd} and then turn over abruptly around 
a column density of $5 \times 10^{19} \cm2$. Below this column 
density these Lyman limit systems must be composed of more diffuse 
ionized gas. 
This is found in hydro simulations \nocite{dhkw:99}({Dav\'{e}} {et~al.} 1999) 
and supported by some observational evidence \nocite{proc:99}({Prochaska} 1999). 
Thus observations at these column densities can directly probe 
whether gas discs reach the values of $N_t$ we require to explain the
DLAS kinematics and $f(N)$. Lastly it is possible to explore how the 
properties of the DLAS evolve with redshift. The merging rate is a 
strong function of redshift \nocite{kola:99,kola:00}({Kolatt} {et~al.} 1999, 2000)
so we would expect the number of 
``multiple hits'' and therefore the kinematics of the DLAS to be 
significantly different at low redshift. All these issues will be 
explored in greater detail in subsequent papers.

If one is really to accept our conclusion that high redshift discs 
have an extended Mestel-type radial profile,
clearly we must develop a theory for their origin.
It is possible that the standard theory
of disc formation is applicable at low redshift, but that some other
process dominates at higher redshift. For example, mergers are far
more common at high redshift, and the gas fractions of discs are
higher (cf. SPF). This may result in efficient transfer of orbital
angular momentum to the gas, producing tidal tails that distribute the
gas out to large radii. Locally, some interacting galaxies show
extremely extended rotating HI, presumably resulting from such a
mechanism \nocite{hibb:99}({Hibbard} 1999). Another possibility is that
starbursts triggered by mergers produce supernovae-driven outflows
like those seen in local starbursts
\nocite{heck:99}({Heckman} 1999) that could also result in extended gas distributions. 
We find that a toy model in which the gas clouds have a bulk velocity
equal to the rotation velocity of the disc, but in a random
direction (i.e. not in a rotationally supported disc) still produces
good agreement with the observed DLAS kinematics because the 
kinematics of our model are dominated by the motions of the multiple 
discs, not the kinematics within these discs.  

It is worth noting that the surface densities of the gas discs in 
our model would be far below the critical value for star formation 
determined by observations \nocite{kenn:89,kenn:98}{Kennicutt} (1989, 1998), implying that 
there may be very little star formation taking place in a `quiescent'
mode. It is interesting that SPF found that a picture in which 
quiescent star formation at high redshift is very inefficient and 
most of the star formation occurs in merger-induced bursts provides 
the best explanation of the high redshift Lyman-break galaxies. 
Even in the extreme case in which quiescent star formation is 
completely switched off, we find that the starburst mode alone can 
easily produce the observed level of star formation at high redshift.
Thus in the high redshift universe interactions between galaxies
seem to play a rather prominent role in determining their gas 
properties and star formation histories.

\section*{Acknowledgements} 
We thank George Blumenthal, James Bullock, Romeel Dav\'{e}, Avishai
Dekel, Martin Haehnelt, Tsafrir Kolatt, Risa Wechsler, and Art Wolfe
for stimulating conversations. We thank Lisa Storrie-Lombardi and 
Art Wolfe for allowing us to use their data prior to publication.
This work was supported by NASA and NSF grants at UCSC. AHM and RSS
also acknowledge support from University Fellowships from the Hebrew 
University, Jerusalem and JXP acknowledges support from a Carnegie 
postdoctoral fellowship. 
The bibliography was produced with Jonathon Baker's Astronat package.

\bibliography{}

\end{document}